\documentclass[aps,prl,twocolumn,showpacs]{revtex4}
\usepackage{graphicx}
\def\s2n{S^{\prime}/N}

\def\ms{$M_{\rm S}$}
\def\ma{$M_{\rm A}$}
\begin{document}
\title{Structure Function Scaling in Compressible Super--Alfv\'{e}nic MHD Turbulence}

\author{Paolo Padoan$^{1}$, Raul Jimenez$^{2}$, {\AA}ke Nordlund$^{3}$ and Stanislav Boldyrev$^{4}$}
\affiliation{$^{1}$Physics Department, University of California, San Diego, La Jolla, CA 92093 \\
$^{2}$Department of Physics and Astronomy, University of Pennsylvania, Philadelphia, PA 19104 \\ 
$^{3}$Copenhagen Astronomical Observatory and Theoretical Astrophysics Center, DK--2100, Copenhagen, Denmark \\
$^{4}$Department of Astronomy and Astrophysics, University of Chicago, 
5640 S. Ellis Avenue, Chicago, IL 60637}

\date{\today}

\begin{abstract}

Supersonic turbulent flows of magnetized gas are believed to play an important role
in the dynamics of star--forming clouds in galaxies. Understanding statistical properties 
of such flows is crucial for developing a theory of star formation. 
In this letter we propose a unified approach for obtaining the velocity scaling in compressible 
and super--Alfv\'{e}nic turbulence, valid for arbitrary sonic Mach number, 
\ms. We demonstrate with numerical simulations that the scaling can be described 
with the She--L\'{e}v\^{e}que formalism, where only one parameter, interpreted
as the Hausdorff dimension of the most intense dissipative structures, needs to be
varied as a function of \ms. Our results thus provide a method for obtaining the velocity
scaling in interstellar clouds once their Mach numbers have been inferred from observations. 

\end{abstract}

\pacs{98.38.Am, 95.30.Qd, 98.38.Dq, 95.30.Lz}

\maketitle

{\bf 1.} {\it Introduction.}
Star formation occurs in our galaxy primarily inside cold interstellar clouds.
Observations indicate that gas motion inside these clouds is highly supersonic 
in the broad range of scales $0.1-1000$~pc. At these scales the clouds lack 
any characteristic structure and their scaling
properties are to some extent universal. This universality is the signature of the
supersonic and super--Alfv\'{e}nic magneto--hydrodynamic (MHD) turbulence that
governs the dynamics of the clouds \cite{Padoan+99per,Ballesteros+99,Padoan+Nordlund99MHD,Klessen+2000,Ostriker+2001,Padoan+2002ext,Padoan+2002scaling}.

The properties of turbulent MHD flows are controlled by the
rms sonic Mach number, \ms, (ratio of rms flow velocity and sound velocity)
and the rms Alfv\'{e}nic Mach number, \ma, (ratio of rms flow velocity and
Alfv\'{e}n velocity). In interstellar clouds these numbers are large,
$M_{\rm S}\sim 1-30$, $M_{\rm A}>1$. The corresponding supersonic motion
produces the observed strong density enhancements \cite{Padoan+2001cores},
some of which are Jeans--unstable and undergo gravitational collapse.
This initial `turbulent' stage of density fragmentation should be described 
statistically, in terms of velocity and density probability distribution
functions or correlators; its properties are inherently related to the 
statistical properties of compressible MHD turbulence \cite{Padoan+Nordlund2002imf}.  

Early studies suggesting the importance of astrophysical supersonic turbulence 
were undertaken by von Weizs\"acker~\cite{von51} and later by Larson~\cite{Larson81}, who discovered 
that the spectrum of the velocity field in molecular clouds was steeper than
the spectrum predicted by the standard Kolmogorov model. The discrepancy was
attributed to the strong intermittency of supersonic turbulence. However, the
analytical understanding of such turbulence remained rather poor. 

Advances in computational and observational techniques have recently allowed
measurements of power spectra and higher order correlators of velocity and density 
fields, reviving the interest in supersonic turbulence. Modern observations generate
surveys covering a range of scales comparable or even exceeding the resolution of
numerical simulations. The interstellar medium can now be viewed as a laboratory
for investigating supersonic motion unachievable with Earth--based experiments.

The turbulent field can be characterized by the scaling of its structure functions
[see below]. The intermittency is defined as the departure of this scaling from the
Kolmogorov one. So far, only the limiting cases of $M_{\rm S}\gg1$ and $M_{\rm S}=0$
have been investigated. An analytical description of supersonic turbulence,
$M_{\rm S}\gg1$, has been proposed in~\cite{Boldyrev2002} and confirmed
numerically in \cite{Boldyrev+2002scaling,Boldyrev+2002structure,Kritsuk+Norman2003}.
It is based on the observation that the most intense dissipative structures in
supersonic flows are two--dimensional shocks. This approach makes use of the model
by She and L\'ev\^eque~\cite{She+Leveque94} for incompressible turbulence, $M_{\rm S}=0$, where the 
dissipative structures are one--dimensional vortex filaments.

In this letter we report the results of a series of numerical MHD simulations 
where we traced the change of the velocity structure functions over the broad
range of Mach numbers $0.3\le M_{\rm S}\le 10$. 
\begin{figure}[ht]
\includegraphics[width=\columnwidth]{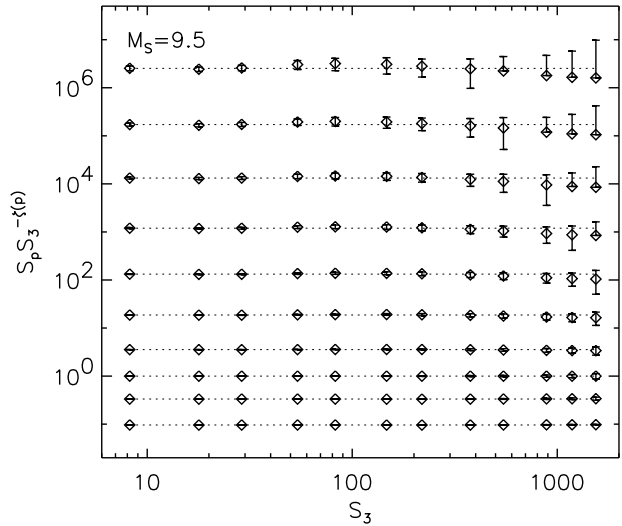}
\caption[]{Compensated velocity structure functions of order $p=1$ 
to $p=10$ (bottom to top) from the \ms=9.5 experiment. To avoid confusion, 
the structure functions of order $p=1$ and $p=2$ have been multiplied by 
0.2 and 0.7 respectively and only the upper part of error bars larger than 
100\% is shown.}
\label{fig1}
\end{figure}
We discovered that their scalings
can be described by the general model where only one parameter, the dimension
of the most intense dissipative structures, needs to be changed with \ms.  
The scalings obtained in \cite{She+Leveque94,Boldyrev2002} naturally appear
as the limiting cases of this unifying model. Our results suggest that
the velocity scaling within interstellar clouds can be inferred from
measurements of their Mach number. Observationally, this is a much easier
task than measuring the structure function directly. We present the main results
here; the detailed discussion will appear elsewhere.

{\bf 2.} {\it Velocity Structure Function Scaling.}
The velocity structure functions of order $p$ are defined as:
\begin{equation}
S_p(\ell)=\langle|u({x}+{\ell})-u({x})|^p\rangle\propto \ell\,^{\zeta(p)}
\label{str}
\end{equation}
where the velocity component $u$ is parallel (longitudinal structure function)
or perpendicular (transversal structure function) to the vector $\ell$
and the spatial average is over all values of the position $x$.

The scaling of velocity structure functions in incompressible
turbulence is best described by the She--L\'{e}v\^{e}que formula 
\cite{She+Leveque94,Dubrulle94}:
\begin{equation}
\zeta(p)/\zeta(3)=p/9 + 2\left[ 1-\left( 2/3\right) ^{p/3}\right
],
\label{s-l}
\end{equation}
This is a specific case of the more general formula
\begin{equation}
\zeta(p)/\zeta(3)=\gamma \, p + C(1-\beta\,^p)
\label{dubrulle}
\end{equation}
with the normalization condition,
\begin{equation}
C=(1-3\gamma)/(1-\beta^3).
\label{cod}
\end{equation}
%
%
\begin{figure}[ht]
\includegraphics[width=\columnwidth]{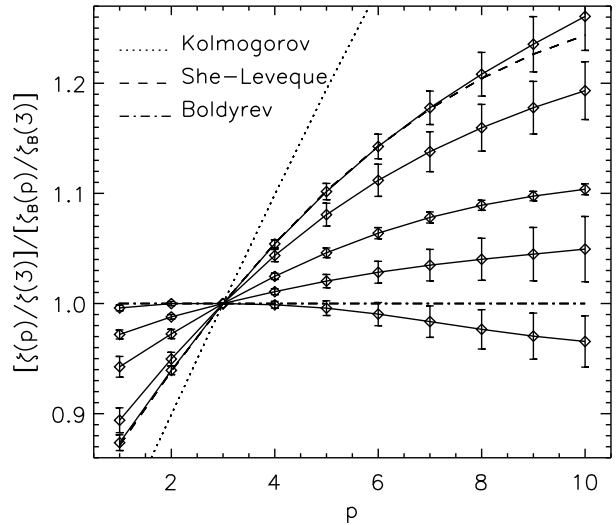}
\caption[]{Structure function exponents normalized to the values
predicted by \cite{Boldyrev2002} for supersonic turbulence 
(dashed--dotted line). Values from five experiments are shown, from \ms=9.5 (bottom) to 
\ms=0.4 (top). The \ms=0.8 case is not shown as it almost overlaps with the 
\ms=0.4 experiment.}
\label{fig2}
\end{figure}
Two basic assumptions are used to derive (\ref{dubrulle}),
as discussed in \cite{She+2001}. The first assumption is the 
existence of a symmetric relation between the intensity of fluctuations 
$F_p(\ell)$ and $F_{p+1}(\ell)$ with respect to a translation in $p$: 
\begin{equation}
F_{p+1}(\ell)=A_p F_p(\ell)^\beta F_{(\infty)}(\ell)^{1-\beta},
\label{scaling}
\end{equation}
where the $p$th order intensity of fluctuations is defined as 
$F_p(\ell)=S_{p+1}(\ell)/S_p(\ell)$,
%
%
and $A_p$ are constants independent of $\ell$ (generally found to be
independent of $p$ as well). The functions $F_p(\ell)$ are associated
with higher intensity fluctuations with increasing values of $p$.
The parameter $\beta$ is a measure of intermittency.
In Kolmogorov's 1941 model \cite{Kolmogorov41}, $\beta \to 1$ and $\zeta(p)=p/3$ 
This is the limit of no intermittency. Lower $\beta$ corresponds to higher 
degree of intermittency, or to the increasing role played by strong 
persistent fluctuations (structures).

The second assumption is
\begin{equation}
F_{\infty}\sim S_3^{\gamma}.
\label{gamma}
\end{equation}
The value of the parameter $\gamma$ is related to very high order moments 
and is therefore difficult to constrain with experiments. Kolmogorov's 
turbulence corresponds to $\gamma=1/3$. 

The parameter $C$ has been interpreted as the Hausdorff codimension of 
the support of the most singular dissipative structures \cite{Dubrulle94}. 
In incompressible turbulence the most dissipative structures are along 
vortex filaments with Hausdorff dimension $D=1$ and so $C=2$. 
\begin{figure}[ht]
\includegraphics[width=\columnwidth]{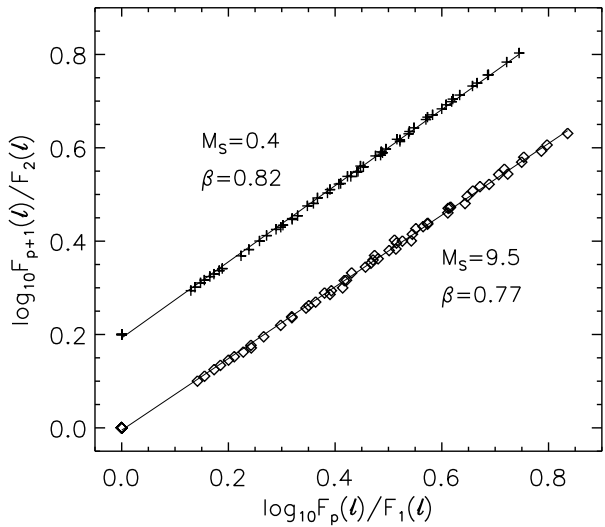}
\caption{``$\beta$ test'' plot for the models with \ms=0.4 and
\ms=9.5. The \ms=0.4 model is shifted upward by 0.2 
y--units. The solid lines are least square fits with slope $\beta$.}
\label{fig3}
\end{figure}
Boldyrev~\cite{Boldyrev2002} has proposed 
to extend the scaling (\ref{dubrulle})
to supersonic turbulence, with the assumption that the
Hausdorff dimension of the support of the most singular dissipative 
structures is $D=2$ ($C=1$), because dissipation of supersonic turbulence
occurs mainly in sheet--like shocks. He used $\gamma=1/9$ as in incompressible
turbulence, which implies $\beta^3=1/3$, that is a higher degree of 
intermittency than in incompressible turbulence. With these parameters
one obtains the velocity scaling proposed for supersonic turbulence
in~\cite{Boldyrev2002}:
\begin{equation}
\zeta(p)/\zeta(3)=p/9 + 1-\left( 1/3\right) ^{p/3}.
\label{boldyrev}
\end{equation}
This velocity scaling has been found to provide a very accurate
prediction for numerical simulations of highly supersonic and 
super--Alfv\'{e}nic turbulence \cite{Boldyrev+2002scaling,Boldyrev+2002structure}.
The same scaling has also been proposed for incompressible 
MHD turbulence \cite{Muller+Biskamp2000}, where dissipation occurs mainly in 
two dimensional current sheets.

In the present work we obtain the structure function scaling 
in simulations of compressible super--Alfv\'{e}nic MHD turbulence as a function
of the rms sonic Mach number of the flow, \ms, for constant value of the rms 
Alfv\'{e}nic Mach number, \ma$\gg 1$ (the limiting case of strong magnetic 
field was investigated analytically and numerically in 
\cite{Lithwick+Goldreich2001,Cho+2002,Cho+Lazarian2002}).
We use an isothermal equation of state and vary the value of \ms\ in different
experiments by varying the thermal energy. As a result, the initial value of
\ma=10 remains unchanged from run to run. In our numerical method the total 
magnetic flux is conserved. However, the magnetic energy is amplified.

The initial value of the ratio of average magnetic and dynamic pressures is
$\langle P_{\rm m} \rangle _{\rm in} / \langle P_{\rm d} \rangle _{\rm in}=0.005$ 
for all runs.

The value of the same ratio, averaged over the time interval for which the structure 
functions are computed, is given in Table~1. Because the average dynamic pressure 
is always in excess of the average magnetic pressure, 
$\langle P_{\rm m} \rangle / \langle P_{\rm d} \rangle \ll 1$, 
the dissipation is expected to occur primarily in vortex 
filaments (roughly corresponding to magnetic flux tubes) in subsonic 
runs and in shocks in supersonic runs. In the next section we argue that
when we increase the value of \ms, the Hausdorff dimension of the support 
of the most dissipative structures grows continuously from $D\approx 1$ to
$D\approx 2$.

\begin{figure}[ht]
\includegraphics[width=\columnwidth]{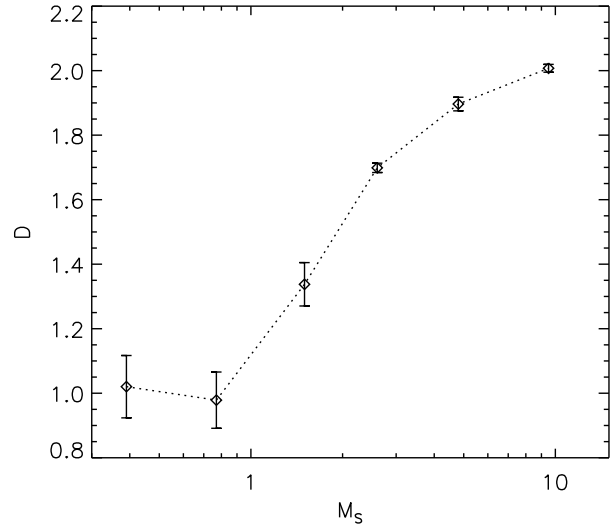}
\caption{Hausdorff dimension of the support of the most dissipative
structures versus rms sonic Mach number.}
\label{fig4}
\end{figure}

{\bf 3.} {\it Results.}
We have solved the three dimensional compressible MHD equations,
for an isothermal gas with initially uniform density and magnetic 
fields, in a staggered mesh with 250$^3$ computational cells and 
periodic boundary conditions. 
Turbulence is set up as an initial large scale random and
solenoidal velocity field (generated in Fourier space with power 
only in the range of wavenumber $1\le K\le 2$) and maintained with 
an external large scale random and solenoidal force, correlated at
the largest scale turn--over time. Details about the
numerical method are given in \cite{Padoan+Nordlund99MHD}. 

We have run six experiments with different values of \ms, 
in the range $0.4\le M_{\rm S}\le9.5$ (Table~\ref{t1}), for
approximately 10 dynamical times. The transverse structure 
functions of velocity are computed for the last four dynamical 
times of each experiment, up to the order $p=10$. Their power law 
slope, $\zeta(p)$, is obtained from a least square fit in the range 
$2\le \ell \le 96$. The standard deviation of the slope is used as 
an estimate of the uncertainty of $\zeta(p)$. 

Figure~\ref{fig1} shows the structure functions plotted versus the 
third order one for the \ms=9.5 experiment. The structure functions are 
compensated by their least square fit slope, $\zeta(p)$. We study the 
structure functions relative to the third order to exploit the concept of 
extended self--similarity \cite{Benzi+93,Dubrulle94}, that is the property 
of the relative scaling to extend up to a range of scales affected by 
dissipation (numerical dissipation in this case). 

The exponents $\zeta(p)$, normalized to the third order one, 
are plotted in Figure~\ref{fig2}, divided by their value predicted 
in~\cite{Boldyrev2002} and given by (\ref{boldyrev}).  
The \ms=0.4 and \ms=0.8 models follow the She--L\'{e}v\^{e}que scaling of 
incompressible turbulence. As the value of \ms\ increases, the velocity 
scaling becomes gradually more intermittent. At \ms=9.5 the scaling is
very close to the analytical prediction for supersonic turbulence 
\cite{Boldyrev2002}, as previously shown 
\cite{Boldyrev+2002scaling,Boldyrev+2002structure}.

The validity of the assumption of hierarchical symmetry expressed by 
(\ref{scaling}), can be tested with a log--log plot
of $F_{p+1}/F_2$ versus $F_p/F_1$, 
\begin{table}[ht]
\begin{tabular}{l|ccccccc}
\hline
\hline
$\langle M_{\rm S}\rangle$  & 0.39 & 0.77 & 1.5 & 2.6  & 4.8  & 9.5  \\
\hline
$\langle P_{\rm m} \rangle _{\rm in} / \langle P_{\rm d} \rangle _{\rm in}$ & 0.005 & 0.005 & 0.005 & 0.005 & 0.005 & 0.005 \\
$\langle P_{\rm m} \rangle / \langle P_{\rm d} \rangle$ & 0.15 & 0.14 & 0.09 & 0.11 & 0.09 & 0.08 \\
$\beta$              & 0.87  & 0.87  & 0.84   & 0.79   & 0.73   & 0.69  \\
$D$                  & 1.0   & 1.0   & 1.34   & 1.70   & 1.90   & 2.01  \\
\hline
\end{tabular}
\caption{Values of the rms sonic Mach number, average magnetic to dynamic pressure
ratio at initial time and averaged over the time interval for which the
structure functions are computed, $\beta$ parameter from fitting
$\zeta(p)/\zeta(3)$ and Hausdorff dimension of the support of the 
most dissipative structures.}
\label{t1}
\end{table}
obtained by varying the argument $\ell$ in (\ref{scaling}). 
If the plot is a straight line, the parameters $A_p$ are constants independent
of $\ell$ and the assumed hierarchical symmetry expressed by (\ref{scaling}) is 
satisfied. This is called the ``$\beta$ test'' in \cite{She+2001}. Our numerical 
simulations pass the $\beta$ test for all values of \ms.
In Figure~\ref{fig3} we show the two extreme cases, \ms=0.4 and \ms=9.5. 
We have used values of $F_p(\ell)$ 
for all values of $\ell$ used in the least square fit estimate
of the exponents $\zeta(p)$. In all the runs we find that 
the constants $A_p$ in (\ref{scaling}) are independent not only of
$\ell$, but also of $p$, as in previous numerical and laboratory studies 
\cite{Leveque+She97,She+2001,Chavarria+95}. The plot in Figure~\ref{fig3}
is in fact obtained varying both $\ell$ and $p$.

The slope of the plot in Figure~\ref{fig3} provides
an estimate of $\beta$. For the models with \ms\ from 0.4 to 
9.5 we obtain $\beta=0.82$, 0.81, 0.75, 0.74, 0.77 and 0.77. 
We instead find no evidence of a systematic variation in the parameter 
$\gamma$ with \ms\ and our data is consistent with $\gamma=1/9$. 

The $\beta$ test gives support to the basic assumption of the
She--L\'{e}v\^{e}que formalism, but does not necessarily yield the 
value of $\beta$ providing the best fit to the structure
function scaling $\zeta(p)/\zeta(3)$. Values of $\beta$ 
computed from a best fit to $\zeta(p)/\zeta(3)$ from our 
numerical experiments are slightly different from
the $\beta$ test values (see Table~\ref{t1}). 
We use these best--fit $\beta$ values to compute
the parameter $D$.

>From the definition of the parameter $C$ in (\ref{cod}), the
values of $\beta$ estimated above and $\gamma=1/9$ we 
can compute $D=3-C$ as a function of \ms. The values of $\beta$ and
$D$ are given in Table~\ref{t1}. The value of $D$
grows with \ms, from $D\approx 1.0$ in the subsonic runs
to $D\approx 2.0$ in the most supersonic runs. 

{\bf 4.} {\it Conclusions.}
We propose a unifying description of the velocity structure
function scaling of compressible and super--Alfv\'{e}nic MHD
turbulence for an arbitrary value of the rms sonic Mach number of the
flow, \ms. Only one model parameter, interpreted as the Hausdorff 
dimension of the support of the most dissipative structures in the 
She--L\'{e}v\^{e}que formalism, needs to be varied as a function
of \ms. Besides its general importance, our result implies that  
the velocity structure function scaling in interstellar 
clouds can be predicted because the value of \ms\ in the clouds is 
easily estimated from observations. This may help improve our understanding
of the process of cloud fragmentation that leads to the birth of stars.\\

\begin{acknowledgments}
RJ was supported by NSF grant AST-0206031. 
{\AA}N was supported by a grant from the Danish
Natural Science Research Council, and in part by the Danish National
Research Foundation, through its establishment of the Theoretical
Astrophysics Center. SB was supported by the NSF Center for magnetic 
self--organization in astrophysical and laboratory plasmas at the University 
of Chicago. 
\end{acknowledgments}


\end{document}